\documentclass{emulateapj}

\usepackage{epsfig}

\slugcomment{astro-ph, 2006 June 27}
\shorttitle{VARIABILITY IN THE LENGTH OF THE SOLAR CYCLE}
\shortauthors{Rogers, Richards \& Richards}

\begin{document}


\title{Long-term Variability in the Length of the Solar Cycle}


\author{Michael L. Rogers, Mercedes T. Richards}

\affil{Department of Astronomy \& Astrophysics, Penn State University,
525 Davey Laboratory, University Park, PA, 16802-6305}

\and
\author{Donald St. P. Richards}

\affil{Department of Statistics, Penn State University,
326 Thomas Building, University Park, PA, 16802-2111}
\email{mtr@astro.psu.edu, mrogers@astro.psu.edu, richards@stat.psu.edu}

\rightskip=0pt 
\begin{abstract}

\rightskip=0pt

Detailed models of the solar cycle require information about the starting time 
and rise time as well as the shape and amplitude of the cycle. However, none of 
these models includes a discussion of the variations in the length of the cycle, 
which has been known to vary from $\sim$7 to 17 years.  The focus of our study 
was to investigate whether this range was associated with a secular pattern in 
the length of the sunspot cycle.  To provide a basis for the analysis of
the long-term behavior of the Sun, we analyzed archival data of sunspot
numbers from 1700 - 2005 and sunspot areas from 1874 - 2005.  The independent
techniques of power spectrum analysis and phase dispersion minimization were
used to confirm the $\sim$11-year Schwabe Cycle, and to illustrate the large 
range in the length of this cycle.  Long-term cycles were identified in archival 
data from 1610 -- 2000 using median trace analyses of the length of the cycle, and 
from power spectrum analyses of the (O-C) residuals of the dates of sunspot minima 
and maxima.  The median trace analysis suggested that the cycle length had a period 
of 183 - 243 years, while the more precise power spectrum analysis identified a 
period of 188 $\pm$ 38 years.  We found that the 188-year cycle was consistent with 
the variation of sunspot numbers and seems to be related to the Schwabe
Cycle.  We found a correlation between the times of historic minima and the
length of the sunspot cycle such that the length of the cycle was usually
highest when the actual number of sunspots was lowest.  The cycle length was
growing during the Maunder Minimum when there were almost no sunspots visible
on the Sun.  This information can now be used to improve the accuracy of the 
current solar cycle models, to better predict the starting time of a given cycle.

\end{abstract}

\vskip10pt


\section{Introduction}

Solar Cycle 23 was predicted to reach a maximum in early 2000
\citep{joselynetal96}, with severe geomagnetic storms expected between 1999
and 2005.  However, this cycle had modest activity compared to the previous
two cycles ({\it e.g.}, \citealt{detomaetal04}).  Our work was motivated by
the publicity associated with this solar cycle because of the relatively high
level of solar activity exhibited through flares in 2003 October and November.
Balasubramaniam \& Regan (1994) showed that the temporal behavior of solar
flares was similar to a sunspot butterfly diagram, with a typical 11-year
cycle.  Our interest in predicting magnetic activity cycles associated with
other cool stars ({\it e.g.}, \citealt{richardsetal03}) led us to investigate
the long-term behavior of the solar cycle.  Earlier analyses of the sunspot
cycle have been summarized by \citet{kuklin76} and \citet{hathaway+wilson04}, 
and a review of the long-term solar cycle was presented by \citet{usoskin+mursula03}.

In 1843, Heinrich Schwabe reported that he had identified a possible 10-year
periodicity in the pattern of sunspots based on his 17-year study of the Sun
from 1826--1843 \citep{schwabe}.  (Note the typographical error in \citet{wilson94} 
that the result was based on a 43-year study.)  In 1848, Rudolph Wolf 
introduced the relative sunspot number (R)
and organized a program of daily observations of sunspots.  He reanalyzed all
earlier data and found that the average length of a solar cycle was about 11 yrs. 
For more than two centuries, solar physicists have applied a variety of techniques
to determine the nature of the solar cycle.  The earliest methods involved counting
sunspot numbers and determining durations of cyclic activity from sunspot minimum
to minimum using the ``smoothed monthly mean sunspot number'' \citep{waldmeier61, wilson87,
wilson94}.  The ``Group sunspot number'' introduced by \citet{hoyt+schatten98} is a
better documented data set than the database of relative sunspot numbers, but the
two data sets since 1882 are virtually identical \citep{hathaway+wilson04}.

Since 1874, sunspot areas have been measured in terms of the total surface area of
the solar disk covered by sunspots at a given time \citep{hath04}.  The analysis of
sunspot numbers or sunspot areas is often referred to as a one-dimensional approach
because there is only one independent variable, namely sunspot numbers or areas
\citep{wilson94}.  Recently, \citet{lietal05} introduced a new parameter called the 
``sunspot unit area'' in an effort to combine the information about the sunspot 
numbers and sunspot areas to derive the length of the cycle.  There is also a 
two-dimensional approach in which the latitude of an observed sunspot is introduced 
as a second independent variable \citep{wilson94}.  When sunspots first appear on 
the solar surface they tend to originate at latitudes around 40 degrees and migrate 
toward the solar equator.  When such migrant activity is taken into account it can 
be shown that there is an overlap between successive cycles, since a new cycle begins 
while its predecessor is still decaying.  This overlap became obvious when 
\citet{maunder04} published his butterfly diagram and demonstrated the latitude drift 
of sunspots throughout the cycles.  Maunder's butterfly diagram showed that although 
the length of time between sunspot minima is on average 11 years, successive cycles
actually overlap by $\sim$ 1 to 2 years.  In addition, \citet{wilson87} found that
there were distinct solar cycles lasting 10 years as well as 12 years.  This type
of behavior suggests that there could be a cyclic pattern in the length of the
sunspot cycle.

Sunspot number data collected prior to the 1700's show epochs in which there were
almost no sunspots visible on the solar surface.  One such epoch, known as the
Maunder Minimum, occurred between the years 1642 and 1705, during which the number
of sunspots recorded were very low in comparison to later epochs \citep{wilson94}.
Other studies include the analysis of geophysical data and tree-ring radiocarbon
data, which contained residual traces of solar activity \citep{bal85}, to determine
if the Maunder period truly had a lower number of sunspots or whether it was simply
a period in which little data were collected or large degrees of errors existed.
While the data collected before the 1700's are typically less reliable than those
collected in more recent times, the timing of the Maunder Minimum is still
relatively accurate because of the correlation with geophysical data.  Other epochs
of minimum solar activity in the past have been noted, such as the Oort Minimum 
from 1010 - 1050, the Wolf Minimum from 1280 - 1340, the Sp\"{o}rer Minimum from 
1420 - 1530 \citep{eddy77, stuiver80, siscoe80}, and the Dalton Minimum from 
1795 - 1823 \citep{soon+yaskell03}.  These minima have been derived from
historical sunspot records, auroral histories \citep{eddy76}, and physical models
linking dendrochronologically dated radiocarbon concentrations to the solar cycle
\citep{solankietal04}.

Currently, the fine details of a given solar cycle can be predicted accurately
only after a cycle has begun ({\it e.g.}, \citealt{elling+schwentek92,
hathawayetal94}; and many others) because many solar cycle models
require information about the starting time and rise time as well as the shape
and amplitude of the cycle.  Most of these models have focussed on the
analysis of the International sunspot numbers of as many previous cycles as
possible in order to predict the pattern for the current cycle.   However, none of 
these models includes a discussion of the variations in the length of the cycle, 
which has been known to vary from $\sim$7 to 17 years.

\begin{deluxetable}{lll}
\tablecolumns{3}
\tablecaption{Duration of the Data}
\tablewidth{0pc}
\tablehead{
\colhead{Data Set} & \colhead{} & \colhead{Duration of Data}}
\startdata
Spot Numbers & Daily & 1818 Jan 8 - 2005 Jan 31 \\
& Monthly & 1749 Jan - 2005 Jan \\
& Yearly & 1700 - 2004 \\
\cline{1-3}
Spot Areas & Daily & 1874 May 9 - 2005 Feb 28 \\
& Monthly & 1874 May - 2005 Feb \\
& Yearly & 1874 - 2004
\enddata
\end{deluxetable}

In this paper, we have investigated  the variations in the length of the
sunspot number cycle and examined whether the variability can be explained 
in terms of a secular pattern.  Toward this goal, we applied classical one-dimensional 
techniques to rederive the periodicities of solar activity using the sunspot number 
and area data to provide internal consistency in our analysis of the long-term 
behavior.  These results were used as a basis in the study of the long-term 
behavior of the cycle length.  In \S2 we discussed the source of the data; in \S3
we described the derivation of the cycle from sunspot numbers and sunspot areas 
using two independent techniques; in \S4 we examined the variability in the cycle 
length based on the times of cycle minima and maxima using two independent 
techniques; and in \S5 we discussed the results.

\vskip20pt
\section{Data Collection}

\begin{figure*}
\epsscale{1}
\epsfig{file=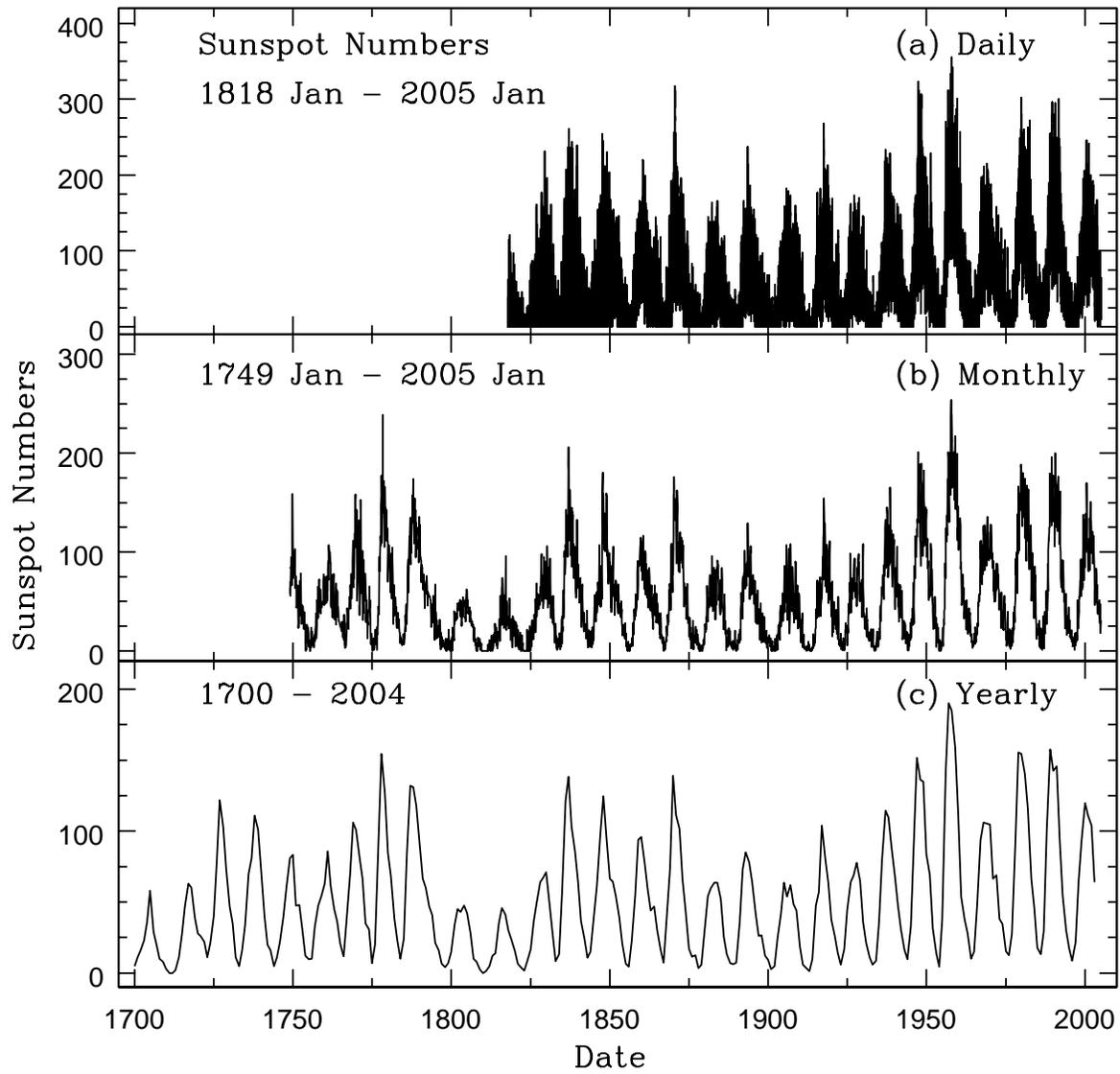,width=16.5cm}
\caption{Archival data for (a) daily, (b) monthly, and (c) yearly sunspot numbers 
from 1700 to 2005. \label{f1}}
\end{figure*}

The sunspot data were collected from archival sources that catalog sunspot numbers,
sunspot areas, as well as the measured length of the sunspot cycle.  The sunspot
number data, ranging from 1700 - 2005, were archived by the National Geophysical
Data Center (NGDC).  These data are listed in individual sets of daily, monthly,
and yearly numbers.  The relative sunspot number, R, is defined as R = K (10g + s),
where g is the number of sunspot groups, s is the total number of distinct spots,
and the scale factor K (usually less than unity) depends on the observer and is
``intended to effect the conversion to the scale originated by Wolf'' \citep{ngdc}.
The scale factor was 1 for the original Wolf sunspot number calculation.  The spot
number data sets are tabulated in Table 1 and plotted in Figure \ref{f1}.

\begin{figure*}
\epsscale{1}
\epsfig{file=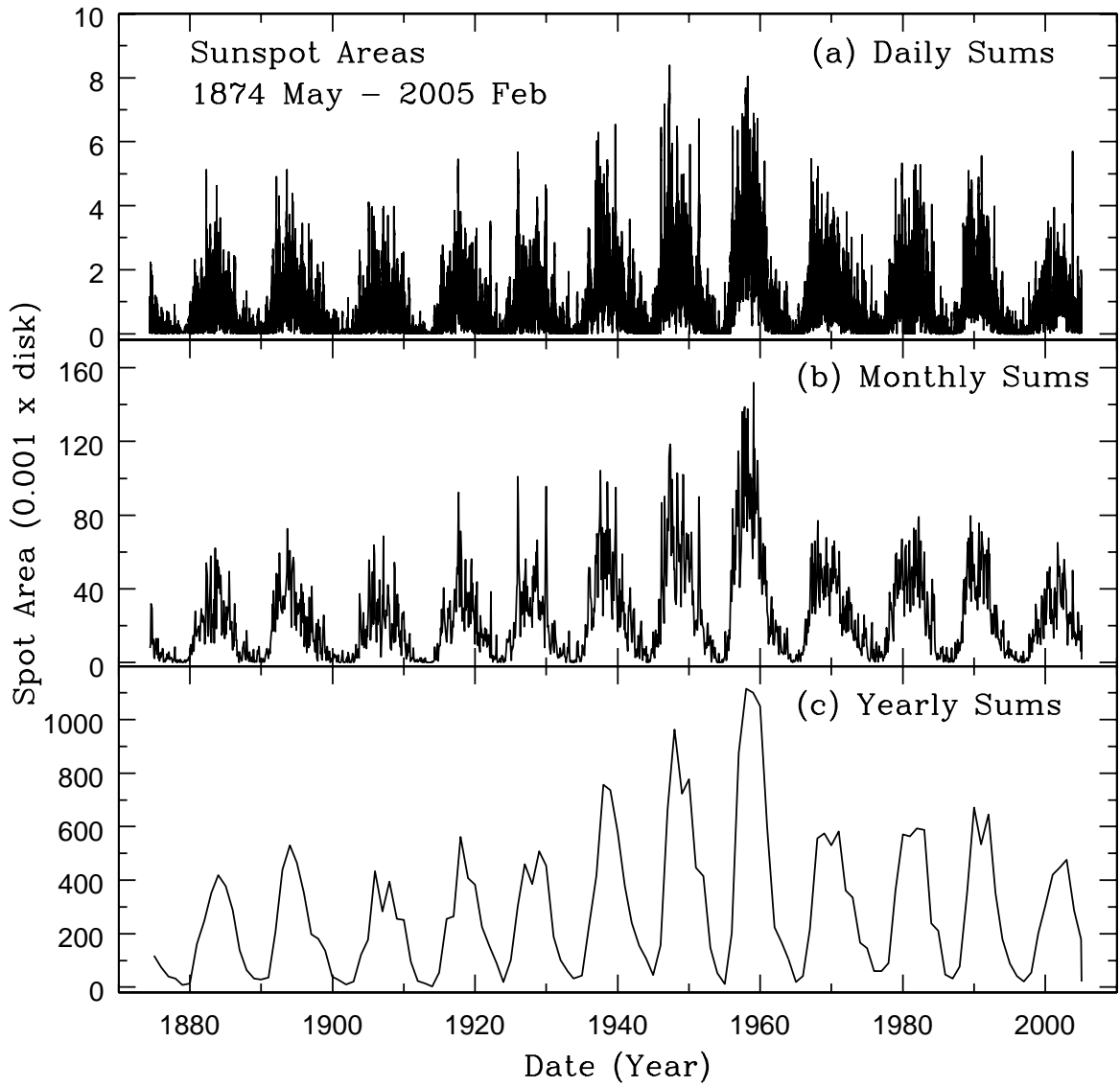,width=16.5cm}
\caption{Archival data for (a) daily, (b) monthly, and (c) yearly sums of 
whole sunspot areas
(0.001 x Solar Hemispheres) from 1874 May to 2005 February. 
\label{f2}}
\end{figure*}

The sunspot area data, beginning in 1874 May 9, were compiled by the Royal
Greenwich Observatory from a small network of observatories \citep{hath04}.  In
1976, the United States Air Force began compiling its own database from its Solar
Optical Observing Network (SOON).  The work continued with the help of the National
Oceanic and Atmospheric Administration (NOAA) \citep{hath04}.  The NASA compilation
of these separate data sets lists sunspot area as the total whole spot area in
millionths of solar hemispheres.  We have analyzed the compiled daily sunspot areas
as well as their monthly and yearly sums.  The sunspot area data sets were tabulated
in Table 1 and plotted in Figure \ref{f2}.  Since the sunspot number and area data
were collected in different ways and by different groups, there may be subtle
inherent differences between the two data sets.  However, these differences should
reveal themselves when the data are analyzed.

The sunspot number cycle data were tabulated by the NGDC \citep{ngdc} and are
discussed further in \S 4.  These data span dates from 1610 to 2000 (Table 2).
Note that the final cycle length was derived from a predicted Cycle 23 minimum
of 1996.5 and a predicted maximum of 2000.3.

{\baselineskip=10pt
\begin{deluxetable}{cccc}
\tablecolumns{4}
\tablewidth{0pc}
\tablecaption{Length of the Sunspot Cycle}
\tablehead{
\colhead{} & \colhead{} & \colhead{Cycle Length} & \colhead{Cycle Length}
\\
\colhead{Year of} & \colhead{Year of} & \colhead{(from minima)} & \colhead{(from maxima)} \\
\colhead{Minimum } & \colhead{Maximum} &  \colhead{(yr)} & \colhead{(yr)}}
\startdata
1610.8 & 1615.5 & ~8.2 & 10.5\\
1619.0 & 1626.0 & 15.0 & 13.5\\
1634.0 & 1639.5 & 11.0 & ~9.5\\
1645.0 & 1649.0 & 10.0 & 11.0\\
1655.0 & 1660.0 & 11.0 & 15.0\\
1666.0 & 1675.0 & 13.5 & 10.0\\
1679.5 & 1685.0 & 10.0 & ~8.0\\
1689.0 & 1693.0 & ~8.5 & 12.5\\
1698.0 & 1705.5 & 14.0 & 12.7\\
1712.0 & 1718.2 & 11.5 & ~9.3\\
1723.5 & 1727.5 & 10.5 & 11.2\\
1734.0 & 1738.7 & 11.0 & 11.6\\
1745.0 & 1750.3 & 10.2 & 11.2\\
1755.2 & 1761.5 & 11.3 & ~8.2\\
1766.5 & 1769.7 & ~9.0 & ~8.7\\
1775.5 & 1778.4 & ~9.2 & ~9.7\\
1784.7 & 1788.1 & 13.6 & 17.1\\
1798.3 & 1805.2 & 12.3 & 11.2\\
1810.6 & 1816.4 & 12.7 & 13.5\\
1823.3 & 1829.9 & 10.6 & ~7.3\\
1833.9 & 1837.2 & ~9.6 & 10.9\\
1843.5 & 1848.1 & 12.5 & 12.0\\
1856.0 & 1860.1 & 11.2 & 10.5\\
1867.2 & 1870.6 & 11.7 & 13.3\\
1878.9 & 1883.9 & 10.7 & 10.2\\
1889.6 & 1894.1 & 12.1 & 12.9\\
1901.7 & 1907.0 & 11.9 & 10.6\\
1913.6 & 1917.6 & 10.0 & 10.8\\
1923.6 & 1928.4 & 10.2 & ~9.0\\
1933.8 & 1937.4 & 10.4 & 10.1\\
1944.2 & 1947.5 & 10.1 & 10.4\\
1954.3 & 1957.9 & 10.6 & 11.0\\
1964.9 & 1968.9 & 11.6 & 11.0\\
1976.5 & 1979.9 & 10.3 & ~9.7\\
1986.8 & 1989.6 & ~9.7 & 10.7\\
1996.5\tablenotemark{1} & 2000.3\tablenotemark{1} & -- & -- \\
\cline{1-4}
Average &  & 11.0$\pm$1.5 & 11.0$\pm$2.0
\enddata
\tablenotetext{1}{Predicted Cycle 23 values from NGDC \citep{ngdc}.}
\end{deluxetable}
}

\vskip20pt
\section{The Length of the Sunspot Cycle from Sunspot Numbers and Areas}

The sunspot number and sunspot area data were analyzed to provide a basis for the
analysis of the long-term behavior of the Sun.  We used the same techniques that
were used by \citet{richardsetal03} in their study of radio flaring cycles of
magnetically active close binary star systems.  

\subsection{Power Spectrum \& PDM Analyses}

Two independent methods were used to determine the solar activity cycles.  In the
first method, we analyzed the power spectrum obtained by calculating the Fast
Fourier transform (FFT) of the data.  The Fourier transform of a function $h(t)$ is
described by $H(\nu) = \int h(t) ~e^{2 \pi i \nu t} dt$ for frequency, $\nu$, and
time, $t$.  This transform becomes a $\delta$ function at frequencies that
correspond to true periodicities in the data, and subsequently the power spectrum
will have a sharp peak at those frequencies.  The Lomb-Scargle periodogram analysis
for unevenly spaced data was used \citep{pressetal92}.

In the second method, called the Phase Dispersion Minimization (PDM) technique
\citep{stellingwerf78}, a test period was chosen and checked to determine if it
corresponded to a true periodicity in the data.  The goodness of fit parameter,
$\Theta$, approaches zero when the test period is close to a true periodicity.  PDM
produces better results than the FFT in the case of non-sinusoidal data.  The
goodness of fit between a test period $\Pi$ and a true period, $P_{true}$ is given
by the statistic, $\Theta = s^2 / \sigma_t^2$ where, the data are divided into $M$
groups or samples, $$\sigma_t^2 = {\sum (x_i - \bar x)^2\over(N - 1)}, \hskip50pt
s^2 = {\sum ( n_j - 1)s_j^2\over(\sum n_j - M)},$$ $s^2$ is the variance of M
samples within the data set, $x_i$ is a data element ($S_{\nu}$), $\bar x$ is the
mean of the data, $N$ is the number of total data points, $n_j$ is number of data
points contained in the sample $M$, and $s_j$ is the variance of the sample $M$.
If $\Pi \neq P_{true}$, then $s^2 = \sigma_t^2$ and $\Theta = 1$.  However, if $\Pi
= P_{true}$, then $\Theta$ $\to$ 0 (or a local minimum).

All solutions from the two techniques were checked for numerical relationships with
(i) the highest frequency of the data (corresponding to the data sampling
interval), (ii) the lowest frequency of the data, $dt$ (corresponding to the
duration or time interval spanned by the data), (iii) the Nyquist frequency, $N/(2
dt)$, and in the case of PDM solutions (iv) the maximum test period assumed.  A
maximum test period of 260 years was chosen for all data sets, except in the case
of the more extensive yearly sunspot number data when a maximum of 350 years was
assumed.  We chose the same maximum test period for the sunspot area analysis for
consistency with the sunspot number analysis, even though these test periods are
longer than the duration of the area data.

\subsection{Results of Power Spectrum and PDM Analyses \label{fftpdmresults}}

The results from the FFT and PDM analyses of sunspot number and sunspot area data
are illustrated in Figures \ref{f3} and \ref{f4}, corresponding to the daily,
monthly, and yearly sunspot numbers and the daily, monthly, and yearly sunspot
areas, respectively.  In these figures, the top frame shows the power spectrum
derived from the FFT analysis, while the bottom frame shows the $\Theta$-statistic
obtained from the PDM analysis.  We specifically used two independent techniques so 
that we could test for consistency and determine the common patterns evident in 
the data.  The fact that the two techniques produced similar results shows that the 
assumptions made in these techniques have minimal influence on the results.  As 
expected, our results confirmed the work done by earlier studies.

The sunspot cycles derived from these results are summarized in Table 3.  The most
significant periodicities corresponding to the 50 highest powers and the 50 lowest
$\theta$ values suggest that the solar cycle derived from sunspot numbers is 10.95
$\pm$ 0.60 years, while the value derived from sunspot area is 10.65 $\pm$ 0.40
years.  The average sunspot cycle from both the number and area data is 10.80 $\pm$
0.50 years.  The strongest peaks in Figures \ref{f3} and \ref{f4} correspond to this
dominant average periodicity over a range from $\sim$7 years up to $\sim$12
years.  A weaker periodicity was also identified from the PDM analysis with an
average period of 21.90 $\pm$ 0.66 years over a range from $\sim$20 -- 24 years.

\begin{deluxetable}{llcc}
\tablecolumns{5}
\tablecaption{Schwabe Solar Cycle Derived from FFT \& PDM Analyses}
\tablewidth{0pt}
\tablehead{
\colhead{} & & \multicolumn{2}{c}{Schwabe Cycle (yrs)} \\
\cline{3-4}
\colhead{Data Set} & & \colhead{FFT} & \colhead{PDM}}
\startdata
Sunspot Numbers & daily & 10.85 $\pm$0.60 & 10.86 $\pm$0.27 \\
              & monthly & 11.01 $\pm$0.68 & 11.02 $\pm$0.68  \\
               & yearly & 10.95 $\pm$0.72 & 11.01 $\pm$0.64 \\
\cline{1-4}
Average (Numbers) &&& 10.95 $\pm$0.60 \\
\cline{1-4}
Sunspot Areas   & daily & 10.67 $\pm$0.44 & 10.67 $\pm$0.42  \\
              & monthly & 10.67 $\pm$0.39 & 10.66 $\pm$0.39  \\
               & yearly & 10.62 $\pm$0.39 & 10.62 $\pm$0.36 \\
\cline{1-4}
Average (Areas) &&& 10.65 $\pm$0.40 
\\
\cline{1-4}
Average (All data) & &  & 10.80 $\pm$0.50
\enddata
\end{deluxetable}

The errors for the FFT and PDM analyses were derived by measuring the Full Width at
Half Maximum (FWHM) of each dominant peak for each data set.  The $1\sigma$ error
is then defined by $\sigma = {\rm FWHM}/2.35$.  The three averages given in Table 3 were
determined by averaging the dominant solutions from the FFT and PDM analyses for
each data set.  The errors in the averages were determined using standard
techniques \citep{bevington69,topping72}.  While the errors for the sunspot area
results are smaller than those for the spot numbers, the area data are actually
less accurate than the sunspot number data because the measurement error in the
areas may be as high as 30\% \citep{hath04}.  The higher errors for the area data
are related to the difficulty in determining a precise spot boundary.

Longer periodicities that could not be eliminated because of relationships with the
duration of the data set or other frequencies related to the data (as described in
\S 3.1), were also identified with durations ranging from $\sim$90 -- 260 years
(Figures \ref{f3} and \ref{f4}).  These long-term periodicities are discussed
further in the following section.

\vskip20pt
\section{Variability in the Length of the Sunspot Cycle from Cycle Minima and Maxima}

The previous analysis of sunspot data provided some evidence of long term cycles in
the sunspot data.  This secular behavior was studied in greater detail through an
analysis of the dates of sunspot minima and maxima from 1610 to 2000 tabulated by
the NGDC \citep{ngdc}.  Since there have been many concerns about the difficulty in  
deriving the exact times of minima, and the even greater complexity in the 
determination of sunspot maxima, we derived our results independently using the cycle 
minima as well as from the cycle maxima.  The sunspot cycle lengths were derived by 
the NGDC from the dates of successive cycle minima.  In addition, we have used their 
tabulated dates of cycle maxima to calculate the corresponding cycle lengths.  These 
cycle lengths are tabulated in Table 2 and plotted in Figure \ref{f5}.  The data in 
Figure 5 show substantial variability over time.

The cycle lengths derived from the dates of sunspot minima and maxima were analyzed
to search for periodicities in the cycle length using two techniques:  (i) a median
trace analysis and (ii) a power spectrum analysis of the `Observed minus
Calculated' or (O-C) residuals.

\subsection{Median Trace Analysis}

A median trace is a plot of the median value of the data contained within a bin of
a chosen width, for all bins in the data set \citep{hoaglin83}.  The median trace 
analysis depends on the choice of an optimal interval width (OIW).  These OIWs, 
$h_n$, were calculated using three statistical methods based on theoretical arguments 
used routinely to estimate the statistical density function of the data.  The first
method defines the OIW as

\begin{equation}
h_{n,1} = \frac{3.49\,{\tilde s}}{n^{1/3}}
\end{equation}

\noindent 
where $n$ is the number of data points and $\tilde s$, a statistically robust
measure of the standard deviation of the data, is defined as

\begin{equation} {\tilde s} = \frac{1}{n}~\sum_{i=1}^{n} \vert x_i - M \vert ~, \end{equation}
where $M$ is the sample median.  The second method defines the OIW as
\begin{equation}
h_{n,2} = 1.66\,{\tilde s}\left(\log_en\over
n\right)^{1/3}.
\end{equation}
\noindent A third definition of the OIW is given by 
\begin{equation}
h_{n,3} = {2 \times IQR\over n^{1/3}} ~ ,
\end{equation}

\noindent 
where $IQR$ is the interquartile range of the data set.  Optimal bin widths were
determined for three data sets corresponding to the cycle lengths derived from the
(i) cycle minima, (ii) cycle maxima, and (iii) the combined minima and maxima data.
Table 4 lists the solutions for the optimal interval widths ($h_{n,1}, h_{n,2},
h_{n,3}$) for each data set.

Since the values of the optimal bin widths ranged from $\sim$60 -- 120 years, we
tested the impact of different bin widths on our results.  This procedure was
limited by the fact that only 35 sunspot number cycles have elapsed since 1610 (see
Table 2).  The data set can be increased to 70 points if we analyze the combined
values of the length of the solar cycle derived from both the sunspot minima and
the sunspot maxima.  Using our derived OIWs as a basis for our analysis, we
calculated median traces for bin widths of 40, 50, 60, 70, 80, and 90 years.  These
are illustrated in Figure \ref{f6}.  The lower bin widths were included to make
maximum use of the limited number of data points, and the higher bin widths were
excluded because, once binned, there would be too few data points to make those
analyses meaningful.

Figure \ref{f6} shows the binned data (median values) and the
sinusoidal fits to the binned data.  The Least Absolute Error Method
\citep{bates88} was used to produce the sinusoidal fits to the median trace in each
frame of the figure.  These sinusoidal fits illustrate the long-term cyclic
behavior in the length of the sunspot number cycle.  The optimal solution was 
determined by identifying the fits that satisfied two criteria: (1) the cycle periods 
deduced from the three data sets should be nearly the same, and (2) the cyclic 
patterns should be in phase for the three data sets.  Table 5 lists the derived 
cycle periods for all three data sets:  the (a) cycle minima, (b) cycle maxima, and
(c) combined minima and maxima data.

\begin{deluxetable}{lccccc}
\tablecolumns{6} 
\tablecaption{Optimal Interval Widths} 
\tablewidth{0pc} 
\tablehead{  
\colhead{} & \colhead{Data} & \colhead{St. Dev.} & 
\multicolumn{3}{c}{Opt. Bin Width (yrs)}\\
\cline{4-6}
\colhead{Data Set}  & \colhead{$n$} & \colhead{$\tilde s$ (yrs)} & \colhead{$h_{n,1}$} 
& \colhead{$h_{n,2}$}  & \colhead{$h_{n,3}$} }
\startdata 
Cycle Minima & 35 & 97.4 & ~103.9 & 75.4 & 116.6\\
Cycle Maxima & 35 & 97.0 & ~103.4 & 75.1 & 115.4\\
Combined & 70 & 97.3 & ~82.4 & 63.5 & 91.8
\enddata
\end{deluxetable}

\subsection{Results of Median Trace Analysis}

The lengths of the sunspot number cycles tabulated by the National Geophysical Data
Center (Table 2 \& Figure \ref{f5}) show that the basic sunspot number cycle is an
average of (11.0 $\pm$ 1.5) years based on the cycle minima and (11.0 $\pm$ 2.0)
years based on the cycle maxima.  This Schwabe Cycle varies over a range from 8 to
15 years if the cycle lengths are derived from the time between successive minima,
while the range increases to 7 to 17 years if the cycle lengths are derived from
successive maxima.  These variations may be significant even though the data in
Figure \ref{f5} show {\it heteroskedasticity}, i.e., variability in the standard
deviation of the data over time.  Although the range in sunspot cycle durations is
large, the cycle length converged to a mean of 11 years, especially after 1818 as
the accuracy of the data became more reliable.  In particular, the sunspot number
cycle lengths from 1610 - 1750 show a wide range of variance while the cycle
durations since 1818 show a much smaller variance (Figure \ref{f5}).  This variance
may be influenced by the difficulty in identifying the dates of cycle minima and
maxima whenever the sunspot activity is relatively low.  Even after the data became
more accurate there was still a significant $\pm$ 1.5-year range about the 11-year
mean.  The range in the length of this cycle suggests that there may be a hidden
longer-term variability in the Schwabe cycle.

Our median trace analysis of the lengths of the sunspot number cycle uncovered
a long-term cycle with a duration between 146 and 419 years (Table 5), if the
data are binned in groups of 40 to 90 years (see \S 4.1).  Since the median
trace analysis is influenced by the bin size of the data, we determined the
optimal bin width based on the goodness of fit between the median trace and
the corresponding sinusoidal fit (see Figure \ref{f6}).  Based on the sunspot
minima (the best data set), the cycle length was 185 years for the 50-yr bin
width, 243 years for the 60-yr bin, 222 years for the 70-yr bin, 393 years for
the 80-year bin, and 299 years for the 90-yr bin; so we found no direct
relationship between the bin size and the resulting periodicity.  Figure
\ref{f6} also shows the median traces for the data and illustrates that the
optimal bin width is in the range of 50 - 60 years because it is only in these
two cases that the sinusoidal fits are in phase and the derived periods are
approximately equal for all three data sets.  The 50-year median trace
predicts a 183-year sunspot number cycle, while the 60-year trace predicts a
243-year cycle.  Since the observations span $\sim$385 years, there is greater
confidence in the 183-year cycle than in the longer one because at least two
cycles have elapsed since 1610.  Similar long-term cycles ranging from 169 to
189 years have been proposed for several decades \citep{kuklin76}.

\begin{deluxetable}{ccccc} 
\tablecolumns{5} 
\tablecaption{Derived Periodicities of Sunspot Number Cycle} 
\tablewidth{0pc} 
\tablehead{  
\colhead{Bin Width} & \multicolumn{4}{c}{Derived Periodicities (yrs)} \\
\cline{2-5} 
\colhead{(yrs)} &  \colhead{Minima} & \colhead{Maxima} & \colhead{Both} & \colhead{Average}}
\startdata 
40 & 157 & 165 & 146 & 156 $\pm$ 10\\
50 & 185 & 182 & 182 & 183 $\pm$ 2\\
60 & 243 & 243 & 243 & 243 \\
70 & 222 & 273 & 304 & 266 $\pm$ 41\\
80 & 393 & 349 & 419 & 387 $\pm$ 35\\
90 & 299 & 299 & 209 & 269 $\pm$ 52
\enddata
\end{deluxetable}

\subsection{Analysis of the (O-C) Data}

The median trace analysis gives us a rough estimate of the long-term sunspot cycle.
However, an alternative method to derive this secular period is to calculate the
power spectrum of the (O-C) variation of the dates corresponding to the (i) cycle
minima, (ii) cycle maxima, and (iii) the combined minima and maxima.

The following procedure was used to calculate the (O-C) residuals for each of the
data sets given above, based only on the dates of minima and maxima listed in Table
2.  First, we defined the cycle number, $\phi$, to be $\phi = (t_i - t_0)/L$, where
$t_i$ are the individual dates of the extrema, and $t_0$ is the start date for each
data set.  Here, L is the average cycle length (10.95 years) derived independently
by the FFT and PDM analyses from the sunspot number data (\S\ref{fftpdmresults}).
The (O-C) residuals were defined to be $$(O-C) = (t_i - t_0) - (N_c\times L)$$
where, $N_c$ is the integer part of $\phi$ and represents the whole number of
cycles that have elapsed since the start date.  The resulting (O-C) pattern was
normalized by subtracting the linear trend in the data.  This trend was found by
fitting a least squares line to the (O-C) data.  The normalized (O-C) data are
shown in Figure \ref{f7} along with the corresponding power spectra.

\subsection{Results of (O-C) Data Analysis}

The power spectra of the (O-C) data in Figure \ref{f7} show that the long term
variation in the sunspot number cycle has a dominant period of $188 \pm 38$ years.
The Gleissberg cycle was also identified in this analysis, with a period of $87 \pm
13$ years.  The solutions for these analyses are illustrated in Figure \ref{f7} and
tabulated in Table 6.  The $1\sigma$ errors were calculated from the FWHM of the
power spectrum peaks, as described in \S3.2.  The sinusoidal fit to the (O-C)
data in Figure \ref{f7} corresponds to the dominant periodicity of 188 years
identified in the power spectra.  Another cycle with a period of $\sim40$ years
was also found.

\vskip20pt
\section{Discussion and Conclusions}

The possible variability in the length of the sunspot cycle was examined
through a study of archival sunspot data from 1610 -- 2005.  In the
preliminary stage of our study, we analyzed archival data of sunspot numbers
from 1700 - 2005 and sunspot areas from 1874 - 2005 using power spectrum
analysis and phase dispersion minimization.  This analysis showed that the
Schwabe Cycle has a duration of (10.80 $\pm$ 0.50) years (Table 3) and that
this cycle typically ranges from $\sim$10 -- 12 years even though the entire 
range is from $\sim$7 -- 17 years.  

The focus of our study was to investigate whether this range was associated
with a secular pattern in the length of the Schwabe cycle.  We used our
derived value for the Schwabe cycle from Table 3 to examine the long-term
behavior of the cycle.  This analysis was based on NGDC data from 1610--2000, a
period of 386 years (using sunspot minima) or 385 years (using sunspot
maxima).  The long-term cycles were identified using median trace analyses of
the length of the cycle and also from power spectrum analyses of the (O-C)
residuals of the dates of sunspot minima and maxima.  We used independent
approaches because of the inherent uncertainties in deriving the exact times
of minima and the even greater complexity in the determination of sunspot
maxima.  Moreover, we derived our results from both the cycle minima and the
cycle maxima.  The fact that we found similar results from the two data sets
suggests that the methods used to determine these cycles (NGDC data) did not
have any significant impact on our results.

\begin{deluxetable}{lcc}
\tablecolumns{3}
\tablecaption{Derived Long Term Solar Cycles}
\tablewidth{0pc}
\tablehead{
\colhead{} & \colhead{Gleissberg} & \colhead{Secular}\\
\colhead{Data Set} & \colhead{(yrs)} & \colhead{(yrs)}}
\startdata
Cycle Minima & 86.8 $\pm$ ~8.8 & 188 $\pm$ 40\\
Cycle Maxima & 86.3 $\pm$ 18.1 & 187 $\pm$ 37\\
Combined & 86.8 $\pm$ 10.7 & 188 $\pm$ 38\\
\cline{1-3}
Average & 86.6 $\pm$ 12.5 & 188 $\pm$ 38
\enddata
\end{deluxetable}

\begin{figure*}
\epsscale{1.0}
\epsfig{file=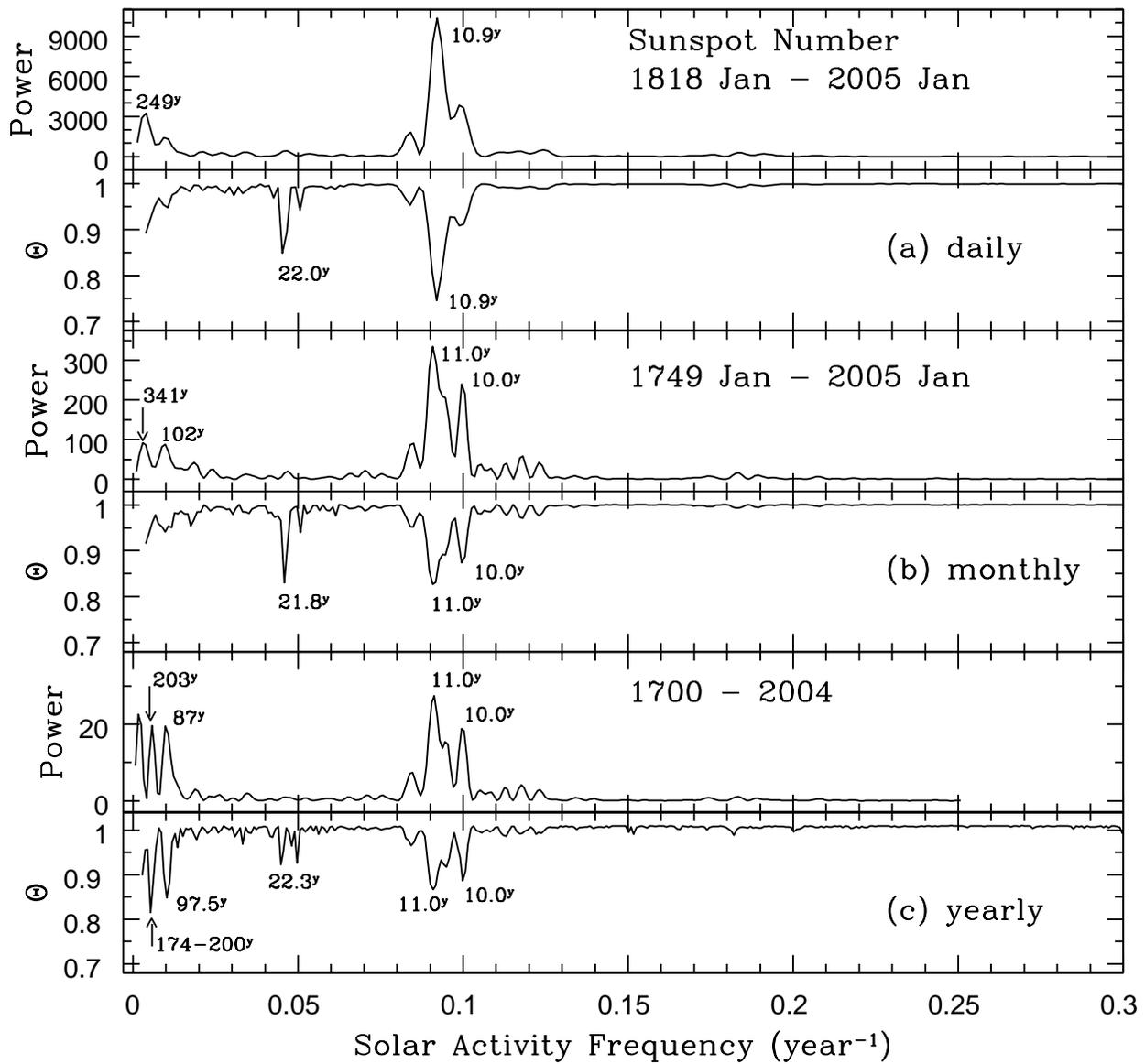,width=17.0cm}
\caption{Frequencies of solar activity derived from power spectrum (upper frame) and PDM
(lower frame) analyses calculated from (a) daily, (b) monthly, and (c) yearly sunspot
numbers.  The labels within the plot show the durations of the derived cycles in units 
of years.
\label{f3}}
\end{figure*}

\begin{figure*}
\epsscale{1.0}
\epsfig{file=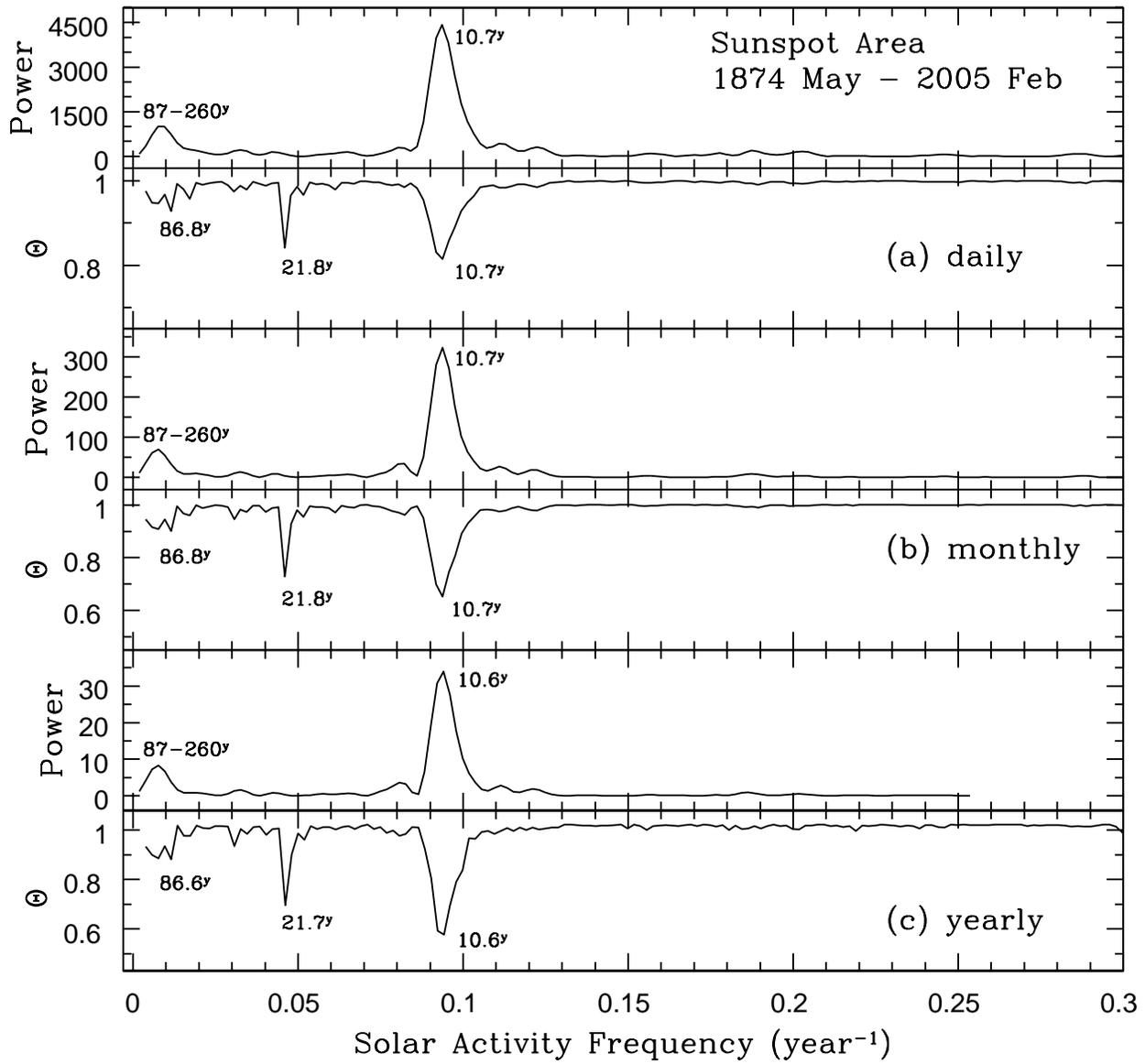,width=17.0cm}
\caption{Frequencies of solar activity derived from power spectrum (upper frame) and PDM
(lower frame) analyses calculated from (a) daily, (b) monthly, and (c) yearly sums of
sunspot areas from 1874 May to 2005 February.  The labels within the plot show the
durations of the derived cycles in units of years. 
\label{f4}}
\end{figure*}

\begin{figure*}
\epsscale{1.0}
\epsfig{file=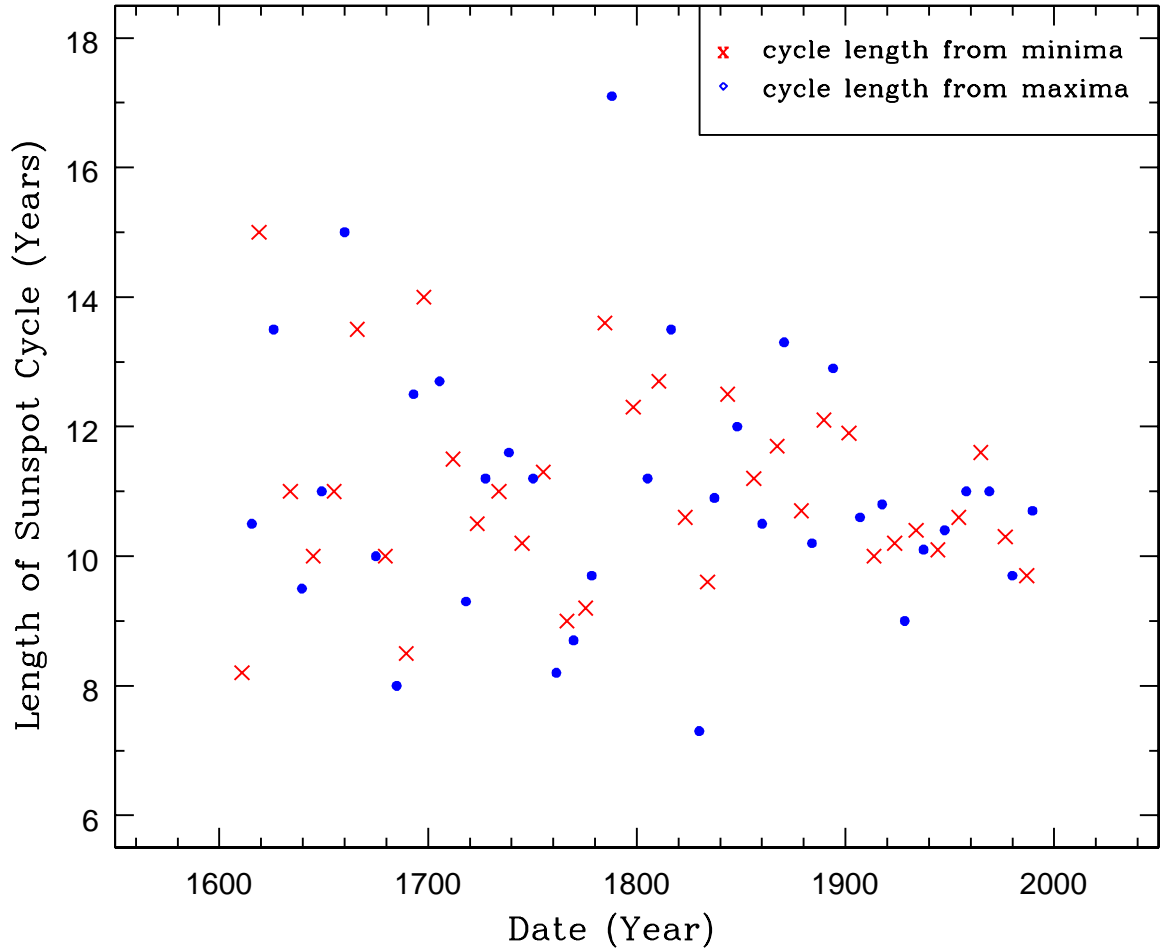,width=17.5cm}
\caption{Sunspot cycle durations derived from successive minima (crosses) and successive maxima
(dots) for dates from 1610.8 to 1989.6. 
\label{f5}}
\end{figure*}

\begin{figure*}
\epsscale{1.0}
\epsfig{file=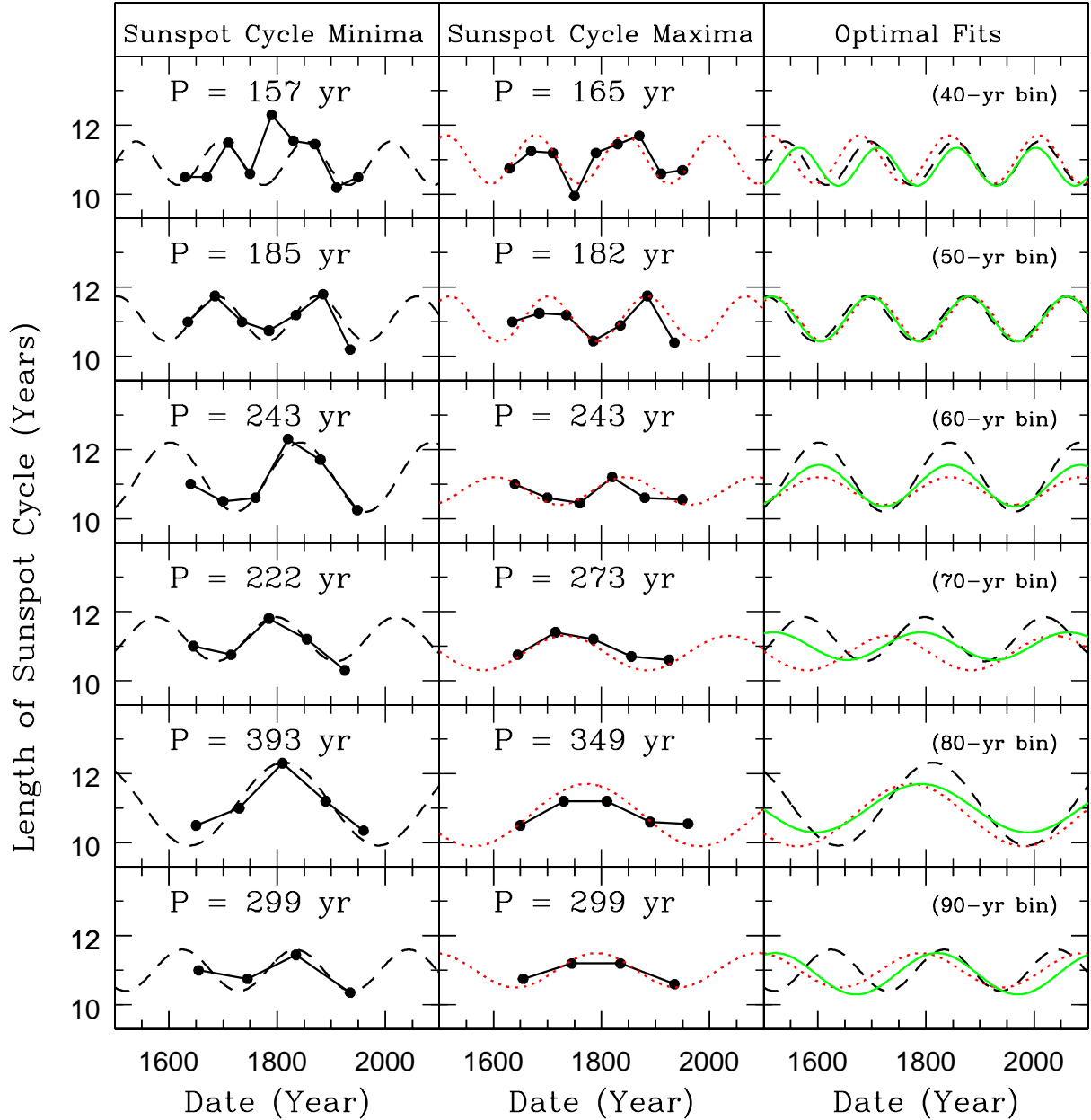,width=17.5cm}
\caption{Median traces for sunspot minima (left column) and maxima (middle column) 
derived for bin widths of 40 -- 90 years.  A sinusoidal fit to the median trace is shown 
for each bin width (minima-dashed line, maxima-dotted line, and combined maxima and 
minima-solid line).  The average period of each derived sinusoidal fit is given at the 
top of each frame.  The optimal fits (right column) show that the optimal
bin width is in the range of 50 -- 60 years because it is only in these two cases that
the sinusoidal fits are in phase and the derived periods are approximately equal for
all three data sets.
\label{f6}}
\end{figure*}

\begin{figure*}
\epsscale{1.0}
\epsfig{file=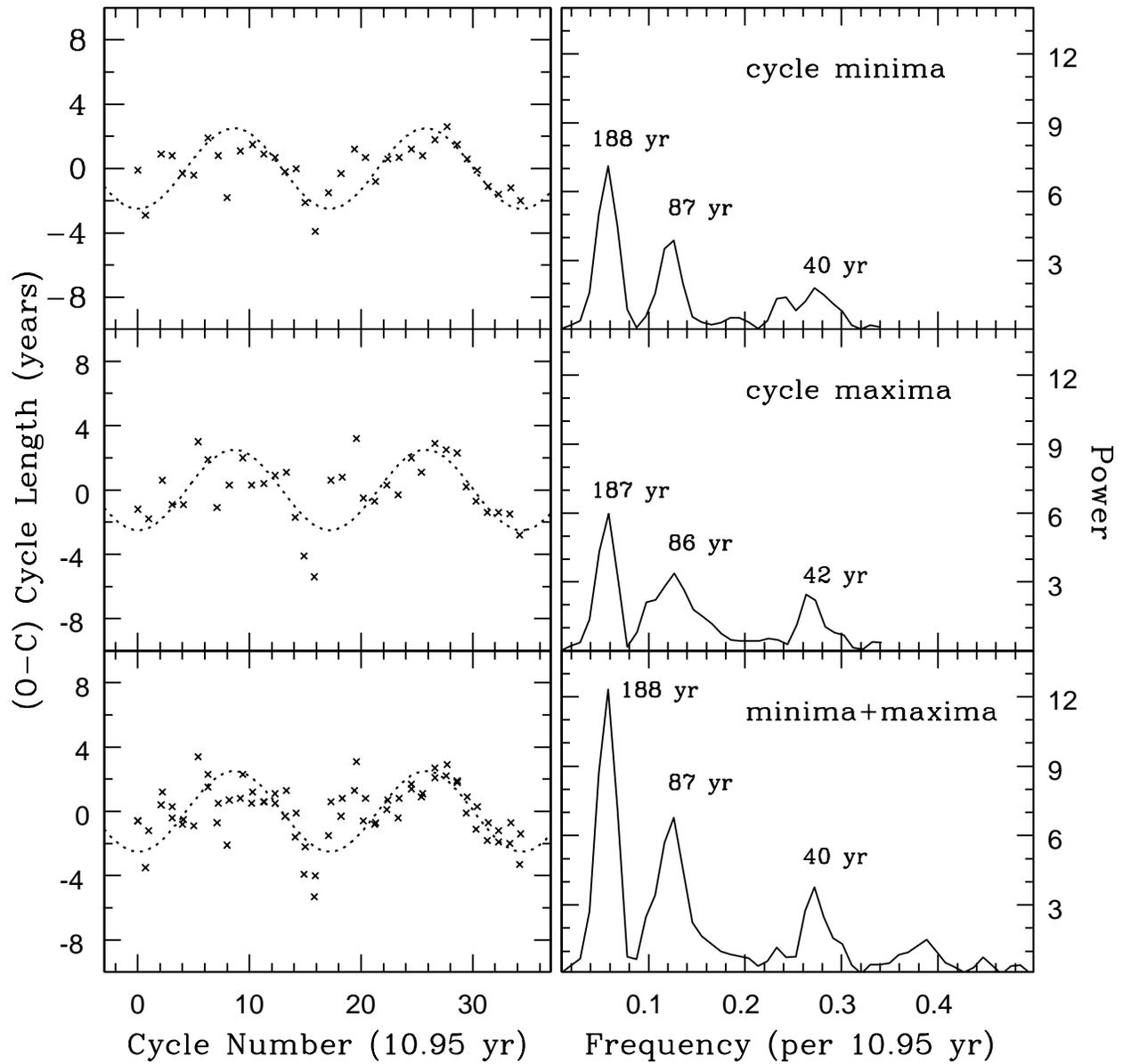,width=17.5cm}
\caption{The cycle length (O-C) residuals (left frames) and the corresponding power 
spectrum (right frames) for the sunspot cycle minima (top frame),
maxima (middle frame), and combined minima and maxima data (bottom frame). The dashed 
line through the data
represents the long term cycle derived from the power spectrum analysis.
\label{f7}}
\end{figure*}

The median trace analysis of the length of the sunspot number cycle provided
secular periodicities of 183 -- 243 years.  This range overlaps with the
long-term cycles of $\sim$90 -- 260 years which were identified directly from
the FFT and PDM analyses of the sunspot number and area data (Figures \ref{f3}
and \ref{f4}).  The power spectrum analysis of the (O-C) residuals of the dates
of minima and maxima provided much clearer evidence of dominant cycles with
periods of $188 \pm 38$ years, $87 \pm 13$ years, and $\sim40$ years.  These
results are significant because at least two long-term cycles have
transpired over the $\sim$385-year duration of the data set.  

\begin{figure*}
\epsscale{1.0}
\epsfig{file=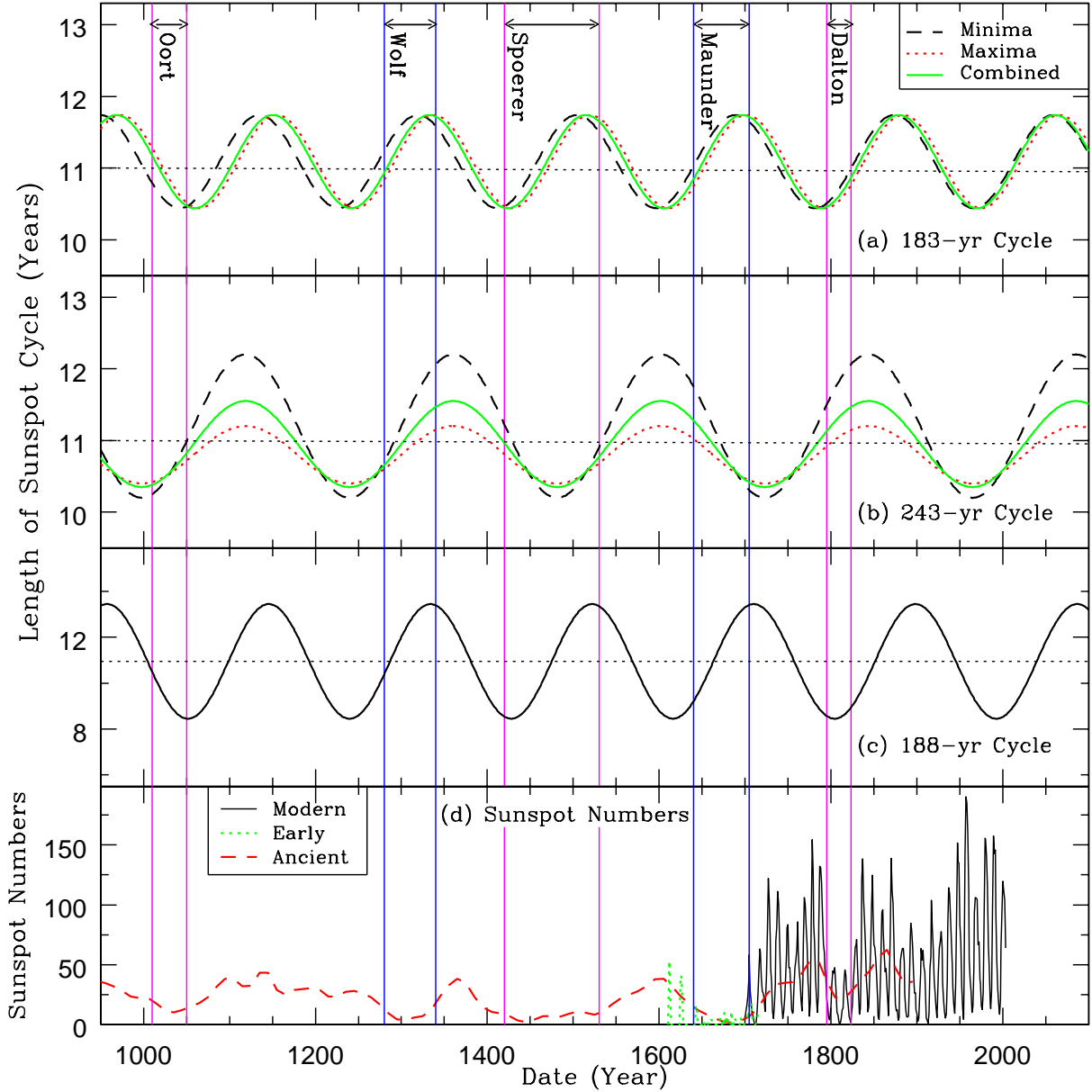,width=16.5cm}
\caption{
Sinusoidal fits to the sunspot number cycle corresponding to the derived periods of (a) 183
years, (b) 243 years, and (c) 188 years, compared with (d) the sunspot number data.  The fits to
(a) and (b) were produced from binned cycle minima (dashed line), cycle maxima (dotted
line), and a combination of the two data sets (solid line).  The bottom frame shows
sunspot numbers from 1700 -- 2004 (modern, solid line), 1610 -- 1715 (early, dotted line), and
950 -- 1950 (ancient, dashed line) reconstructed from radiocarbon data.  The Dalton,
Maunder, Sp\"{o}rer, Wolf, and Oort Minima are identified on the graph.  The dates of
these historical minima correspond well with the derived periodicities of 183
and 188 years, except for the Oort Minimum.  The 243-year cycle does not match the
historical minima as well as the other periodicities.
\label{f8}}
\end{figure*}

The derived long-term cycles were compared in Figure \ref{f8} with documented
epochs of significant declines in sunspot activity, like the Oort, Wolf, Sp\"{o}rer,
Maunder, and Dalton Minima \citep{eddy77, stuiver80, siscoe80}.  In
this figure, the modern sunspot number data were combined with earlier data
from 1610-1715 \citep{eddy76} and with reconstructed (ancient) data spanning
the past 11,000 years \citep{solankietal04}.  These reconstructed sunspot
numbers were based on dendrochronologically-dated radiocarbon concentrations
which were derived from models connecting the radiocarbon concentration with
sunspot number \citep{solankietal04}.  The reconstructed sunspot numbers are
consistent with the occurrences of the historical minima (e.g., Maunder
Minimum).  \citet{solankietal04} found that over the past 70 years, the level
of solar activity has been exceptionally strong.  Our 188-year periodicity is
similar to the 205-year de Vries-Seuss cycle which has been
identified from studies of the carbon-14 record derived from tree rings ({\it e.g.},
\citealt{wagneretal01,braunetal05}).

Figure \ref{f8} compares the historical and modern sunspot numbers with the derived
secular cycles of length (a) 183 years (\S4.2), (b) 243-years (\S4.2), and (c) 188
years (\S4.4).  The first two periodicities were derived from the median trace
analysis, while the third one was derived from the power spectrum analysis of the
sunspot number cycle (O-C) residuals.  The fits for the 183-year periodicity all
had the same amplitude, but were moderately out of phase with each other, while the
fits for the 243-year periodicity were perfectly in phase for all data sets, but
with different amplitudes.  This figure shows that the 183- and 188-year cycles
provided a more consistent match to the sunspot number data than the 243-year cycle,
especially during the Wolf, Sp\"{o}rer, Maunder, and Dalton Minima.  In particular,
we note that the more accurately determined 188-year cycle is consistent with the
behavior of sunspot numbers over time.  Moreover, the four historic minima since
1200, all occurred during the rising phase of our derived 188-year sunspot cycle
when the length of the sunspot cycle was increasing.  Figure \ref{f8} shows that,
on average, the length of the sunspot cycle was highest when the actual number of
sunspots was lowest.  According to our model, the length of the sunspot cycle was
growing during the Maunder Minimum when there were almost no sunspots visible.

The existence of long-term solar cycles with periods between 90 and 200 years
is not new to the literature.  However, the nature of these cycles is still
not understood.  While \citet{hath04} noted evidence of a cycle of $\sim$90
years in sunspot cycle amplitude variations, our studies of the length of the
sunspot cycle have shown that the dominant cycle in these amplitude variations
is 188 years, with weaker periods of $\sim$40 and 87 years.  Our results
suggest that this 188-year cycle is related to the basic Schwabe Cycle.
While we would like to confirm this result using a longer baseline (beyond 385
years), it should be noted that Schwabe's 10-year period was derived from only
17 years (less than two cycles) of observations \citep{schwabe}.  In addition, it
is well-known from Gauss' work on orbits that an orbit can be uniquely defined
if at least three positions along that orbit are known, as long as those
positions cover critical parts of the orbit.  So, our 188-year result may yet
stand the test of time.  Our model predicts that the length of the sunspot
number cycle should increase gradually over the next $\sim$75 years,
accompanied by a gradual decrease in the number of sunspots. This information can now 
be used to improve the accuracy of the current solar cycle models ({\it e.g.}, 
\citealt{dikpatietal06}) to better predict the starting time of a given cycle.

\vskip5pt
\acknowledgements

We thank K. S. Balasubramaniam for his comments on the manuscript, Alon
Retter for his comments on the research and for his advice on the (O-C) analysis,
and David Heckman for advice on the data analysis.  The SuperMongo plotting program
\citep{lupton77} was used in this research.  This work was partially supported by
Penn State University, NSF grant AST-0074586, and an REU supplement to NSF grant
AST-0434234 (PI - G.J. Babu) awarded to the Center for Astrostatistics at Penn 
State University.


\clearpage

\clearpage

\clearpage

\clearpage

\clearpage
\end{document}